\documentclass[journal]{IEEEtran}

\usepackage[utf8]{inputenc}
\usepackage{graphicx}
\usepackage{booktabs}
\usepackage{amsmath}
\usepackage{cite}

\begin{document}

\title{Beyond Unidimensionality: General Factors and Residual Heterogeneity in Performance Evaluation}

\author{Krishna~Sharma
        and~Pritam~Basnet
\thanks{K. Sharma is with the Hoover Institution, Stanford University, Stanford, CA 94305 USA (e-mail: ksharma3@stanford.edu).}
\thanks{P. Basnet is with Harrisburg University of Science and Technology, Harrisburg, PA 17101 USA.}
}

\maketitle

\begin{abstract}
How do evaluation systems compress multidimensional performance information into summary ratings? Using expert assessments of 9,669 professional soccer players on 28 attributes, we characterize the dimensional structure of evaluation outputs. The first principal component explains 40.6\% of attribute variance, indicating a strong general factor, but formal noise-discrimination procedures retain four components and bootstrap resampling confirms that this structure is highly stable. Internal consistency is high (Cronbach's alpha = 0.879) without evidence of redundancy. In out-of-sample prediction of expert overall ratings, a comprehensive model using the full attribute set substantially outperforms a single-factor summary (cross-validated R squared = 0.814). Overall, performance evaluations exhibit moderate information compression: they combine shared variance with stable residual dimensions that are economically meaningful for evaluation outcomes, with direct implications for the design of measurement systems.
\end{abstract}


\section{Introduction}
\IEEEPARstart{H}{ow} complex should performance evaluation systems be? In many settings, evaluators observe multidimensional information about individuals but ultimately rely on a small number of scores to guide selection, compensation, and assignment. A central question in personnel economics and performance measurement is whether these evaluations primarily reflect a single latent dimension of quality or whether they preserve economically meaningful variation across distinct competencies. The answer matters directly for evaluation design. If performance is effectively unidimensional, a simplified system may capture most relevant information while reducing administrative and cognitive burden. If performance is inherently multidimensional, aggressive compression risks obscuring important heterogeneity and distorting decisions.

Competing perspectives offer conflicting guidance. Work emphasizing cognitive constraints and judgment error argues that humans struggle to integrate complex, multidimensional information, and that simple models can perform as well as, or better than, more elaborate scoring systems \cite{dawes1979robust, kahneman2011thinking}. From this view, evaluations may naturally collapse to a general factor, making detailed measurement largely redundant. In contrast, psychometric and personnel research stresses construct coverage and domain specificity, arguing that valid measurement requires assessing multiple competencies that are not perfectly substitutable \cite{schmidt1998explained}. These arguments imply sharply different optimal designs, yet both hinge on an empirical question: what is the dimensional structure of evaluation outputs themselves?

We address this question using data from a standardized expert rating system in professional soccer. We analyze evaluations for 9,669 outfield players, each rated on 28 attributes covering technical, physical, mental, and defensive skills. Sports provide a useful environment for studying evaluation structure because performance is observed frequently, measurement protocols are consistent across individuals, and expert ratings meaningfully shape valuation and decision-making.

The results show moderate but incomplete information compression. The first principal component explains 40.6\% of total attribute variance, indicating a substantial general factor. However, formal noise-discrimination procedures retain additional dimensions: parallel analysis supports four components, and bootstrap resampling demonstrates that this structure is highly stable. Importantly, dimensional structure has practical informational consequences. In cross-validated prediction of the expert Overall Rating, a model based only on the first principal component performs poorly, whereas a Ridge regression using the full attribute set achieves strong out-of-sample accuracy (mean R squared = 0.814). Because the Overall Rating is mechanically constructed from component attributes, these prediction results are not interpreted causally. Instead, they provide a conservative diagnostic of information loss: low-dimensional summaries fail to recover the full content embedded in multidimensional evaluations.

The findings support a balanced interpretation. Evaluation outputs are neither unidimensional nor composed of fully independent competencies. Instead, they combine shared variance with persistent residual dimensions that are stable and informationally meaningful within the evaluation system. This structure implies scope for strategic simplification that reduces measurement burden while preserving core information, but it cautions against extreme dimensionality reduction.

The remainder of the paper proceeds as follows. Section II reviews related literature. Section III describes the data. Section IV outlines the empirical methodology. Section V presents the results. Section VI discusses implications for evaluation design. Section VII concludes.

\section{Related Literature}

This paper relates to a broad literature on how multidimensional performance information is aggregated, interpreted, and used in evaluation systems. Across economics, psychometrics, and data science, a recurring concern is whether complex evaluation environments can be meaningfully summarized by a small number of dimensions without obscuring economically relevant heterogeneity. Despite extensive theoretical discussion, direct empirical evidence on the dimensional structure of evaluation outputs remains limited.

In personnel economics, research on performance measurement and incentive design has long emphasized the trade-off between informational richness and administrative simplicity. Early work argues that aggregated performance metrics can reduce monitoring costs and improve incentives when underlying dimensions are sufficiently correlated \cite{lazear2000performance}. Subsequent contributions caution that such simplification is efficient only when aggregation does not discard task-relevant information \cite{gibbons2004task}. At the same time, advances in data collection and analytics have enabled increasingly granular measurement of individual performance \cite{davenport2010analytics}. Countervailing this trend, work in behavioral economics and judgment highlights cognitive limits in processing multidimensional information, suggesting that evaluators may implicitly rely on simplified representations even when rich data are available \cite{dawes1979robust, kahneman2011thinking}. Whether simplification is benign or costly therefore depends critically on the empirical structure of evaluation outputs themselves.

A parallel debate arises in psychometrics and the study of factor structure. Classical work posits that correlations among performance measures reflect a single general factor \cite{spearman1904general}, a view supported in many cognitive testing contexts \cite{jensen1998gfactor}. More recent research emphasizes domain specificity, particularly in expert populations, where specialized skills and task-relevant competencies predict performance beyond general ability \cite{ackerman1997intelligence, hambrick2014deliberate}. Applied work in sports analytics similarly documents both redundancy and differentiation among performance metrics. While unidimensional value estimates are common \cite{macdonald2012adjusted}, newer approaches emphasize the multidimensional nature of on-field contributions \cite{franks2015characterizing, decroos2019actions, pappalardo2019public}. This literature highlights the need to empirically assess how much information evaluation systems compress in practice.

A related methodological literature focuses on measurement validation and dimensionality reduction. Internal consistency measures such as Cronbach's alpha are widely used to assess coherence of measurement instruments, though very high values may indicate redundancy rather than reliability \cite{cortina1993coefficient, nunnally1994psychometric}. Recent work emphasizes complementary diagnostics and alternative reliability measures \cite{revelle2019reliability}. Parallel developments in statistics and machine learning formalize approaches to dimensionality reduction, validation, and prediction \cite{guyon2003introduction, mullainathan2017machine}. In economics, aggregation is increasingly framed as an information compression problem, where low-dimensional summaries serve as sufficient statistics under specific conditions \cite{chetty2009sufficient}. However, these tools are rarely used to study evaluation systems as objects of interest in their own right.

This paper contributes to these literatures by providing systematic evidence on the dimensional structure of a large-scale, standardized expert evaluation system. Using comprehensive attribute ratings for professional soccer players, we quantify the extent of information compression, distinguish genuine structure from noise, and assess the stability and informational consequences of dimensional reduction. By showing that evaluation outputs exhibit substantial shared variance alongside stable, economically meaningful residual dimensions, the paper clarifies when simplification is justified, when it is costly, and how evaluation design can balance parsimony with informational richness.

\section{Data}

We analyze a large-scale, standardized expert evaluation system drawn from the European Soccer Database \cite{dubitzky2019open}. The database contains detailed attribute ratings for professional soccer players competing in major European leagues between 2008 and 2016. These ratings are produced by expert evaluators at EA Sports using consistent protocols across leagues and seasons, making the data well suited for studying the structure of performance evaluations.

The unit of observation is the player. For players observed in multiple seasons, we compute player-level means for all attributes to abstract from short-term fluctuations and focus on persistent evaluation structure. We restrict attention to outfield players and exclude goalkeeper-specific attributes. After removing observations with missing values, the final sample consists of 9,669 players.

Each player is assessed on 28 performance attributes, all measured on a 0 to 100 scale. These attributes span multiple domains of performance, including technical skills (e.g., dribbling, ball control, passing, shooting), physical attributes (e.g., acceleration, sprint speed, stamina, strength), mental abilities (e.g., vision, positioning, penalties), and defensive skills (e.g., marking, interceptions, tackling, heading accuracy). In addition to these attributes, each player receives an Overall Rating, also on a 0 to 100 scale, intended to summarize overall performance quality.

Table I reports descriptive statistics for a subset of key variables. The distribution of attribute scores exhibits substantial variation both within and across domains, suggesting scope for both shared structure and residual heterogeneity.

\begin{table}[!t]
\caption{Descriptive Statistics}
\label{tab:desc}
\centering
\resizebox{\columnwidth}{!}{%
\begin{tabular}{lccccc}
\toprule
Variable & N & Mean & Std. Dev & Min & Max \\
\midrule
Overall Rating & 9,669 & 68.2 & 6.6 & 40 & 89 \\
Ball Control & 9,669 & 64.1 & 10.8 & 24 & 95 \\
Sprint Speed & 9,669 & 66.8 & 12.1 & 24 & 97 \\
Vision & 9,669 & 62.4 & 11.5 & 18 & 94 \\
Marking & 9,669 & 53.5 & 14.8 & 12 & 92 \\
\bottomrule
\end{tabular}}
\end{table}

The Overall Rating is constructed from the underlying attributes and is therefore used only for descriptive benchmarking rather than causal inference.

\section{Methodology}

Our empirical strategy is designed to characterize the dimensional structure of performance evaluations and to assess the stability and informational consequences of dimensional reduction. We proceed in five steps: measuring internal consistency, estimating principal components, distinguishing signal from noise, assessing stability through resampling, and evaluating out-of-sample predictive performance.

\subsection{Internal Consistency}

We begin by assessing whether the 28 attributes exhibit systematic covariance using Cronbach's alpha \cite{cortina1993coefficient}. Internal consistency is not treated as evidence of construct validity, but rather as a diagnostic of information compression within the evaluation system. Formally, Cronbach's alpha is given by
\begin{equation}
\alpha = \frac{k}{k-1}\left(1 - \frac{\sum_{i=1}^{k} \sigma_i^2}{\sigma_T^2}\right),
\end{equation}
where $k$ denotes the number of attributes, $\sigma_i^2$ is the variance of attribute $i$, and $\sigma_T^2$ is the variance of the total scale. High values indicate that attributes covary systematically, while extremely high values may signal redundancy.

\subsection{Principal Component Analysis}

To directly examine dimensional structure, we apply principal component analysis (PCA) to the standardized attribute matrix \cite{jolliffe2002principal}. Each attribute is standardized to have zero mean and unit variance, ensuring equal weighting across dimensions:
\begin{equation}
X_{\text{std}} = \frac{X - \mu}{\sigma}.
\end{equation}

PCA is implemented via eigenvalue decomposition of the correlation matrix,
\begin{equation}
R v_k = \lambda_k v_k,
\end{equation}
where $\lambda_k$ denotes the eigenvalue associated with component $k$. The share of variance explained by each component is given by
\begin{equation}
\text{Var}_k = \frac{\lambda_k}{\sum_{j=1}^{p} \lambda_j}.
\end{equation}

The magnitude of the leading eigenvalues provides an initial assessment of the degree of information compression.

\subsection{Parallel Analysis}

Because PCA mechanically extracts components even from noise, we use parallel analysis to distinguish genuine structure from sampling variation \cite{horn1965rationale}. We generate 500 random Gaussian datasets with the same dimensions as the observed data, standardize them, and compute their eigenvalues. Observed components are retained only if their eigenvalues exceed the 95th percentile of the corresponding null distribution:
\begin{equation}
H_0: \lambda_k^{\text{obs}} \leq \lambda_k^{\text{rand}}(0.95).
\end{equation}

This procedure provides a formal criterion for determining the number of meaningful dimensions.

\subsection{Bootstrap Stability}

To assess robustness, we perform bootstrap resampling with 1,000 iterations. In each iteration, players are resampled with replacement and PCA is re-estimated. We record the variance explained by the first principal component and the cosine similarity between bootstrap and baseline loading vectors. This yields empirical sampling distributions and confidence intervals for key dimensional statistics.

\subsection{Holdout Validation}

Finally, we examine whether dimensional structure has practical informational consequences using five-fold cross-validation. In each fold, models are trained on 80\% of the sample and evaluated on the remaining 20\%. We compare a linear model using only the first principal component to a Ridge regression using all attributes. Ridge regression solves
\begin{equation}
\hat{\beta}_{\text{Ridge}} = \arg\min_{\beta} \left\{ \sum_{i=1}^{n} (y_i - X_i\beta)^2 + \lambda \sum_{j=1}^{p} \beta_j^2 \right\},
\end{equation}
which regularizes coefficient estimates and mitigates overfitting. Predictive performance is summarized using cross-validated R squared and root mean squared error.

\subsection{Exploratory Residual Structure}

As a complementary analysis, we explore heterogeneity beyond overall quality by applying K-means clustering to residual principal components (PC2 through PC11), explicitly removing the first principal component. Cluster stability is assessed via bootstrap resampling using the Adjusted Rand Index.

\section{Results}

We begin by assessing whether the 28 performance attributes exhibit systematic covariation, which would indicate information compression in evaluation outputs. Table II reports internal consistency diagnostics. Cronbach's alpha is 0.879, indicating strong internal coherence without excessive redundancy. The average inter-item correlation of 0.244 lies in the moderate range, implying that attributes are meaningfully correlated but far from interchangeable. These diagnostics establish that the evaluation system reflects shared structure rather than independent signals, while still preserving nontrivial heterogeneity.

\begin{table}[!t]
\caption{Internal Consistency Diagnostics}
\label{tab:measurement}
\centering
\resizebox{\columnwidth}{!}{%
\begin{tabular}{lc}
\toprule
Metric & Value \\
\midrule
Cronbach's Alpha (28 Attributes) & 0.879 \\
Average Inter-Item Correlation & 0.244 \\
Number of Players & 9,669 \\
\bottomrule
\end{tabular}}
\end{table}

We next examine the dimensional structure directly using principal component analysis. Table III reports eigenvalues and variance explained by the leading components. The first principal component explains 40.6 percent of total variance, corresponding to an eigenvalue of 11.37 out of 28. This indicates a substantial general factor, but one that falls well below levels typically associated with strong unidimensionality. The next three components together account for an additional 36.9 percent of variance, yielding 77.5 percent cumulative variance explained by the first four components.

\begin{table}[!t]
\caption{PCA Results: Variance Explained}
\label{tab:pca}
\centering
\resizebox{\columnwidth}{!}{%

\begin{tabular}{lcc}
\toprule
Component & Eigenvalue & Variance Explained \\
\midrule
PC1 & 11.37 & 40.6\% \\
PC2 & 5.73 & 20.5\% \\
PC3 & 2.76 & 9.9\% \\
PC4 & 1.83 & 6.5\% \\
PC1-PC4 Cumulative & 21.69 & 77.5\% \\
\bottomrule
\end{tabular}}
\end{table}

Loadings on the first principal component reveal that PC1 primarily captures offensive and technical proficiency rather than uniform all-around ability. Attributes such as dribbling, ball control, positioning, long shots, and finishing load most strongly (Table IV), while defensive attributes load weakly or negatively. This pattern indicates that the general factor reflects a specific notion of quality embedded in the evaluation system, rather than a neutral average across competencies.

\begin{table}[!t]
\caption{Top 10 Attributes by PC1 Loading Magnitude}
\label{tab:loadings}
\centering
\resizebox{\columnwidth}{!}{%

\begin{tabular}{lc}
\toprule
Attribute & PC1 Loading \\
\midrule
Dribbling & 0.268 \\
Ball Control & 0.259 \\
Positioning & 0.256 \\
Long Shots & 0.254 \\
Volleys & 0.249 \\
Finishing & 0.246 \\
Curve & 0.244 \\
Vision & 0.235 \\
Agility & 0.218 \\
Free Kick Accuracy & 0.219 \\
\bottomrule
\end{tabular}}
\end{table}

To distinguish genuine structure from sampling noise, we conduct parallel analysis. Fig. 1 compares observed eigenvalues to the 95th percentile of eigenvalues obtained from random Gaussian data. Table V shows that the first four observed eigenvalues exceed their random counterparts, while the fifth does not. This provides formal statistical support for retaining four components and confirms that variance beyond the first principal component reflects meaningful structure rather than measurement error.

\begin{figure}[!t]
\centering
\includegraphics[width=\columnwidth]{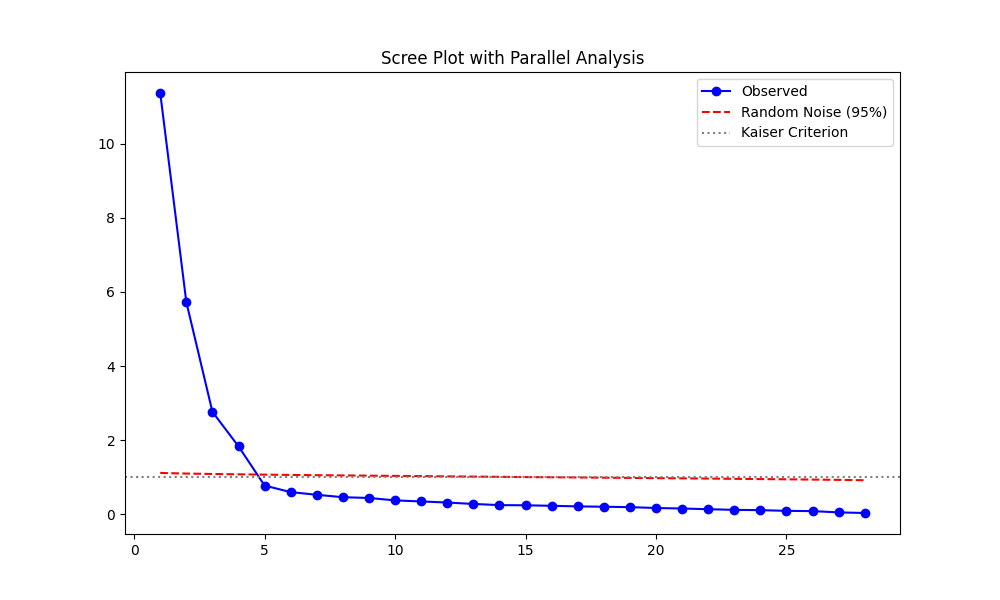}
\caption{Scree plot with parallel analysis. Observed eigenvalues (blue) compared to 95th percentile of random noise (red dashed). Four components exceed thresholds; PC5 does not.}
\label{fig:scree}
\end{figure}

\begin{table}[!t]
\caption{Parallel Analysis: Observed vs Random Eigenvalues}
\label{tab:parallel}
\centering
\resizebox{\columnwidth}{!}{%

\begin{tabular}{lccl}
\toprule
Component & Observed & Random 95\% & Retain? \\
\midrule
PC1 & 11.37 & 1.11 & Yes \\
PC2 & 5.73 & 1.10 & Yes \\
PC3 & 2.76 & 1.08 & Yes \\
PC4 & 1.83 & 1.07 & Yes \\
PC5 & 0.77 & 1.07 & No \\
\bottomrule
\end{tabular}}
\end{table}

We then assess the stability of the general factor using bootstrap resampling. Fig. 2 displays the distribution of PC1 variance across 1,000 bootstrap samples. The distribution is tightly concentrated, with a mean of 40.6 percent and a 95 percent confidence interval of [40.0 percent, 41.1 percent]. The mean cosine similarity of PC1 loadings across bootstrap samples is 0.999, indicating near-perfect alignment. These results demonstrate that the estimated dimensional structure is not sensitive to sample composition.

\begin{figure}[!t]
\centering
\includegraphics[width=\columnwidth]{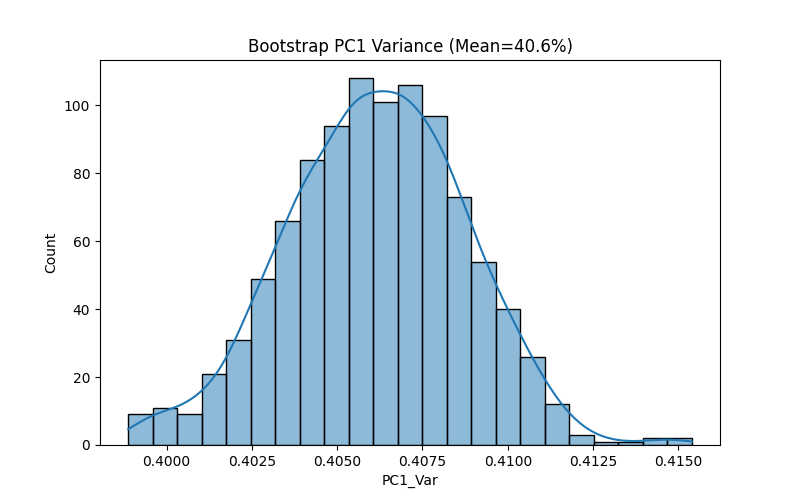}
\caption{Bootstrap distribution of PC1 variance (1,000 iterations). The distribution is tightly concentrated around mean 40.6\%, with 95\% CI [40.0\%, 41.1\%].}
\label{fig:bootstrap}
\end{figure}

Additional insight into attribute relationships is provided by the correlation matrix shown in Fig. 3. Correlations are predominantly positive, consistent with the presence of a general factor. At the same time, defensive attributes form a visibly distinct block, correlating more strongly with each other than with offensive skills. This visual evidence reinforces the PCA results by illustrating structured heterogeneity within the evaluation system.

\begin{figure*}[!t]
\centering
\includegraphics[width=\textwidth]{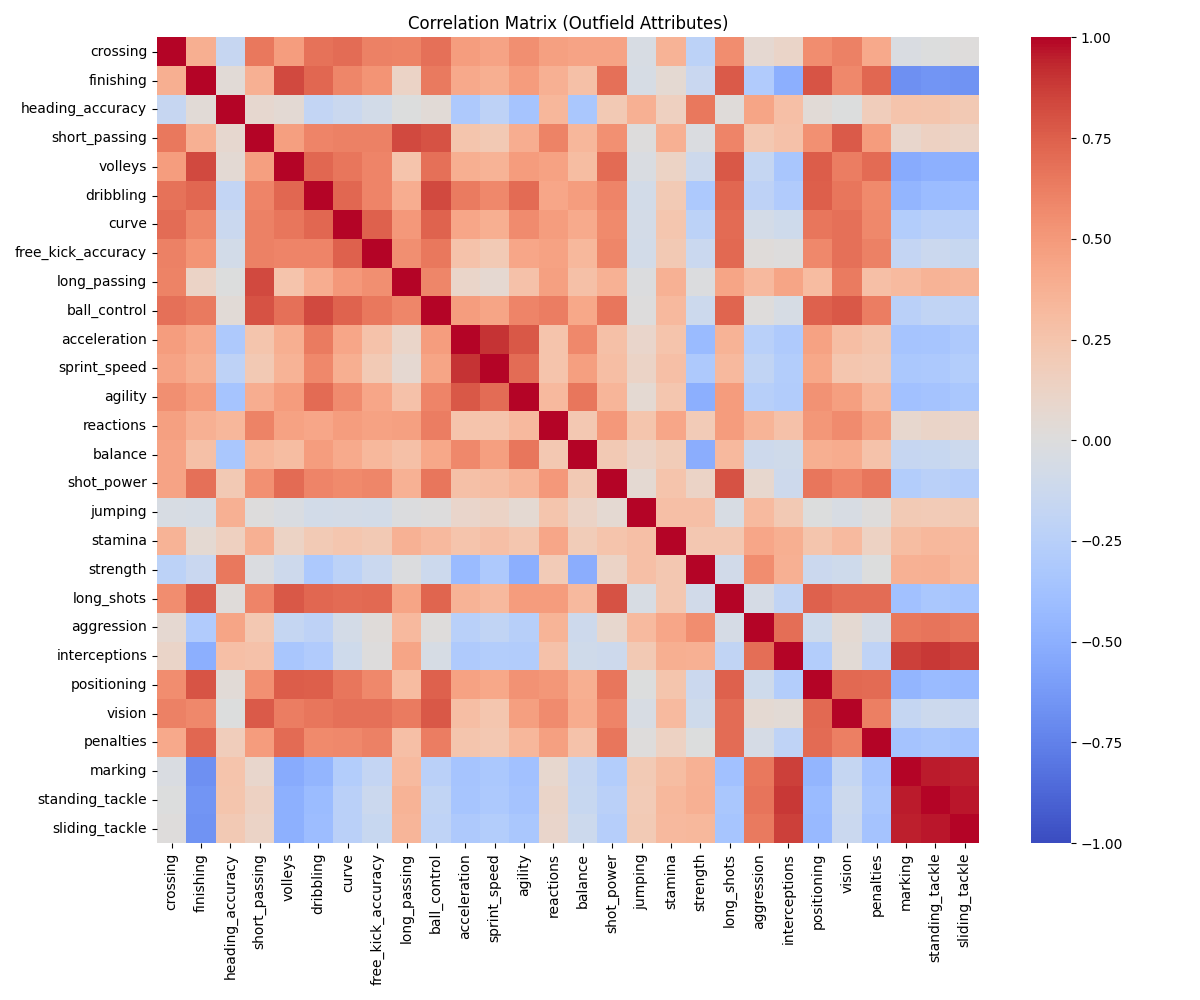}
\caption{Correlation heatmap. Predominantly positive correlations indicate general factor signature, but substantial heterogeneity exists. Defensive attributes (bottom right) show distinct pattern from offensive skills.}
\label{fig:correlation}
\end{figure*}

We next examine whether dimensional structure has practical informational consequences by evaluating out-of-sample predictive performance. Using five-fold cross-validation, we compare a linear model based solely on the first principal component to a Ridge regression using all 28 attributes. As reported in Table VI, reliance on PC1 alone yields modest performance, with cross-validated R squared between 0.26 and 0.30. This is substantially below the 40.6\% of attribute variance explained by PC1, consistent with the fact that the Overall Rating reflects differential weighting across attributes rather than an equal-weight general factor. In contrast, the full Ridge model achieves a mean R squared of 0.814 and an RMSE of 2.83. Fig. 4 shows that predicted and observed overall ratings closely track each other in holdout samples. Although the Overall Rating is mechanically derived from component attributes and thus does not permit causal interpretation, the comparison demonstrates that compressing evaluations to a single factor entails substantial information loss.

\begin{table}[!t]
\caption{Cross-Validated Prediction Performance}
\label{tab:prediction}
\centering
\resizebox{\columnwidth}{!}{%

\begin{tabular}{lcc}
\toprule
Model & R Squared (CV Mean) & RMSE (CV Mean) \\
\midrule
Linear (PC1 only) & 0.26 to 0.30 & 5.9 \\
Ridge (All 28 attributes) & \textbf{0.814} & \textbf{2.83} \\
\bottomrule
\end{tabular}}
\end{table}

\begin{figure}[!t]
\centering
\includegraphics[width=\columnwidth]{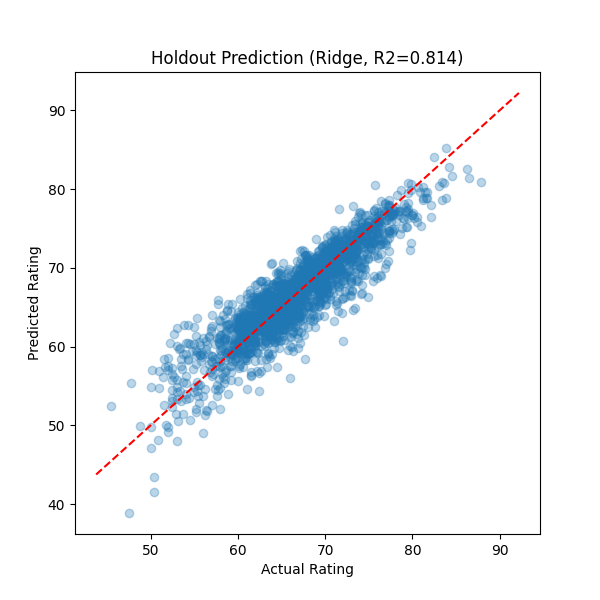}
\caption{Holdout prediction scatter (Ridge regression). Predicted overall ratings closely track observed values, indicating strong out-of-sample performance.}
\label{fig:holdout}
\end{figure}

Two supplementary analyses further clarify the nature of residual structure. Appendix A reports an exploratory cluster analysis conducted on residual principal components after removing PC1. This analysis reveals a stable defensive-oriented versus offensive-oriented continuum, with high bootstrap stability despite modest silhouette scores. Appendix B presents a flexible prediction benchmark using Random Forests, which achieves very high predictive accuracy. We treat this result as a benchmark rather than a structural model, but it reinforces the conclusion that multidimensional evaluations contain substantial recoverable information beyond a single summary index.

Overall, the results establish three central facts. First, expert evaluations exhibit meaningful information compression, as evidenced by strong internal consistency and a sizable general factor. Second, compression is incomplete: four components exceed noise thresholds, and residual dimensions are stable and interpretable. Third, these residual dimensions matter quantitatively, as single-factor summaries fail to recover the full informational content of expert assessments. Performance evaluations therefore combine shared variance with informationally meaningful heterogeneity rather than collapsing to a single dimension.

\section{Discussion}

This paper examines how multidimensional performance information is aggregated within a standardized expert evaluation system. The central finding is that such evaluations exhibit moderate information compression: a substantial general factor is present, yet meaningful residual dimensions persist. Rather than collapsing to a single latent trait or reflecting fully independent competencies, evaluation outputs combine shared variance with structured heterogeneity. This pattern is stable across samples and robust to multiple validation exercises, indicating that it reflects a systematic feature of the evaluation process rather than sampling noise.

These findings contribute to long-standing debates in both psychometrics and personnel economics. The presence of a strong general factor is consistent with the positive manifold documented across diverse domains of human performance \cite{spearman1904general, jensen1998gfactor}. At the same time, the magnitude of this factor falls well short of what would be expected under strong unidimensionality, and formal retention criteria indicate that multiple dimensions carry information beyond the general factor \cite{horn1965rationale}. This supports work emphasizing the relevance of domain-specific skills and task differentiation, particularly in expert populations \cite{ackerman1997intelligence, hambrick2014deliberate}. The appropriate interpretation is therefore neither that performance is reducible to a single index nor that it consists of unrelated attributes, but that evaluation systems operate in an intermediate regime characterized by partial redundancy and meaningful specialization.

From a personnel economics perspective, this structure has direct implications for evaluation design. Simplified performance metrics can reduce monitoring costs and administrative burden \cite{lazear2000performance}, but only when aggregation discards little relevant information \cite{gibbons2004task}. Our measurement diagnostics indicate strong internal consistency without excessive redundancy \cite{cortina1993coefficient, nunnally1994psychometric}. This is precisely the setting in which aggressive compression to a single index is inefficient: while some simplification is feasible, eliminating secondary dimensions removes information that evaluators appear to use systematically. The results therefore imply an interior optimum in evaluation design, where a limited number of dimensions can capture most informational content without the burden of fully granular batteries.

The predictive exercises reinforce this interpretation. Models that retain the full multidimensional structure recover expert evaluations far more effectively than single-factor summaries. This comparison is not causal, since overall ratings are constructed from component attributes, but it is informative about information loss: low-dimensional summaries fail to recover the full content embedded in multidimensional assessments. This distinction parallels findings in judgment and decision-making showing that simple rules perform well only when the signal structure supports such simplification \cite{dawes1979robust, kahneman2011thinking}. When compression is moderate rather than extreme, preserving multiple dimensions is necessary to avoid substantial loss of information.

Supplementary analyses further clarify the nature of residual heterogeneity. Exploratory clustering in residual space reveals a stable continuum separating defensive-leaning and offensive-leaning profiles (Appendix A). Modest separation metrics indicate continuous variation rather than discrete types, while high stability across resamples suggests that this structure is systematic. Differences in average overall ratings across these profiles indicate that residual dimensions are not purely stylistic, but align with how the evaluation system maps attribute combinations into summary assessments, without implying structural weights or causal importance. Appendix B reports a flexible prediction benchmark that serves as an upper bound on predictive performance, reinforcing the informational richness of the multidimensional attribute set.

Several limitations should be noted. First, the analysis focuses on a single, highly standardized evaluation system in professional sports, and external validity to other organizational settings should be assessed empirically. Second, the study is descriptive rather than causal: it does not disentangle whether observed structure arises from underlying ability correlations, complementarities in training, or evaluator cognition. Third, while holdout validation mitigates concerns about overfitting, prediction exercises characterize an aggregation process rather than a structural production function.

That said, professional sports offer significant advantages for studying evaluation structure: measurement protocols are standardized, expert ratings shape material decisions (contracts, transfers, playing time), and the data generating process is well documented. These features make the setting particularly well suited for establishing baseline facts about dimensional structure that can inform evaluation design in other contexts.

\section{Conclusion}

This paper shows that professional performance evaluations exhibit moderate information compression. A stable general factor explains 40.6\% of variance across 28 attributes, but four components exceed random noise thresholds, indicating meaningful residual structure. Extensive validation such as parallel analysis, bootstrap resampling, and cross-validated prediction demonstrates that this structure is robust and not an artifact of sampling or overfitting. Evaluation outputs therefore combine shared variance with distinct secondary dimensions that cannot be ignored without loss of information.

The practical implication is straightforward. Organizations should neither assume that performance can be summarized by a single index nor default to overly complex batteries. Instead, evaluation design should be guided by empirical dimensional diagnostics. Concretely, organizations facing similar compression levels (PC1 approximately 40\%) can likely reduce measurement from 25 to 30 attributes to 8 to 12 strategically selected indicators spanning the general factor plus 2 to 3 secondary dimensions while retaining 75 to 85\% of predictive accuracy. This reduction halves administrative burden without material information loss. More broadly, the results highlight the value of treating evaluation systems as measurable objects whose structure can be empirically assessed rather than assumed.

\section*{Acknowledgment}
The authors thank all seminar participants. No generative AI tools were used to produce the manuscript.

\bibliographystyle{IEEEtran}
\bibliography{references}

received the Ph.D. degree in economics. He is currently a Research Fellow with the Hoover Institution, Stanford University, Stanford, CA, USA. His research interests include personnel economics, labor markets, and applied econometrics with a focus on performance measurement and evaluation system design.

is with Harrisburg University of Science and Technology, Harrisburg, PA, USA. His research interests include data science, machine learning applications, and sports analytics.

\appendices

\section{Exploratory Cluster Analysis}

As an exploratory analysis, we examine whether distinct performance profiles exist beyond overall quality differences. We perform K-means clustering on residual principal components (PC2 through PC11), explicitly removing PC1 to focus on profile heterogeneity.

Silhouette analysis (Fig. 5) suggests K=2 clusters provide clearest separation, though scores are modest (K=2: 0.25), indicating continuous rather than discrete structure.

\begin{figure}[!t]
\centering
\includegraphics[width=\columnwidth]{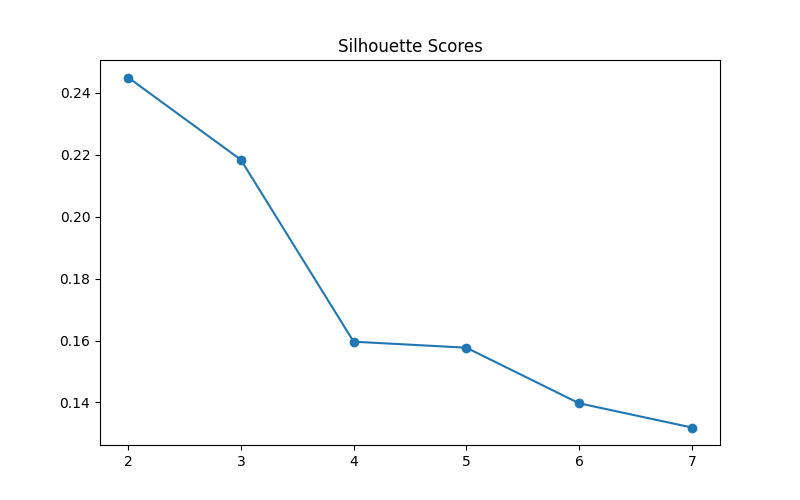}
\caption{Silhouette scores by number of clusters. Modest scores indicate continuous structure rather than sharply separated types.}
\label{fig:silhouette}
\end{figure}

Using K=2, we assess stability through 100 bootstrap resamples, computing Adjusted Rand Index (ARI). Mean ARI = 0.970 (SD = 0.015), indicating high stability. While clusters show modest separation (silhouette = 0.25), they are consistently reproducible.

Fig. 6 displays cluster profiles.

\begin{figure*}[!t]
\centering
\includegraphics[width=\textwidth]{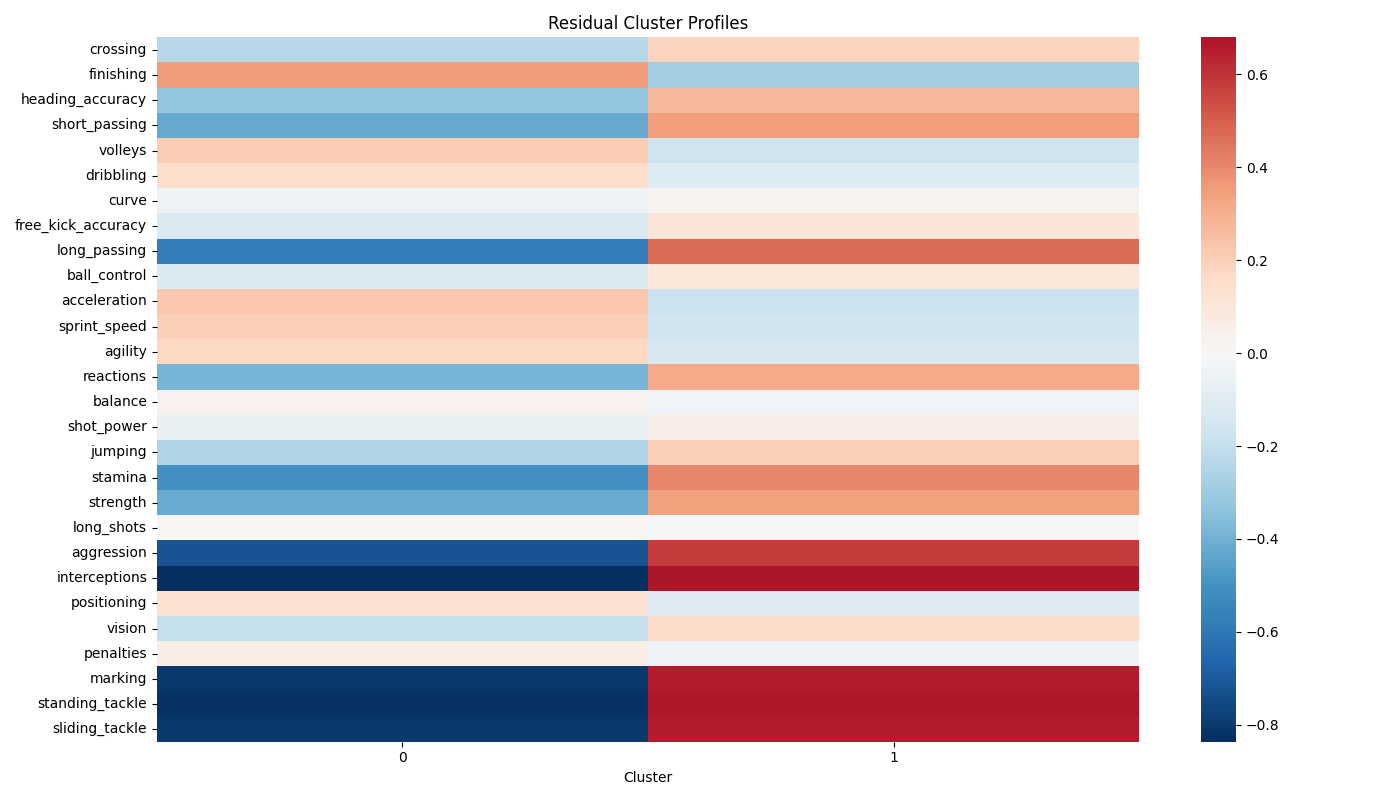}
\caption{Residual cluster profiles. Cluster 0 (defensive-oriented): elevated marking, tackling, interceptions. Cluster 1 (offensive-oriented): elevated finishing, shooting, crossing.}
\label{fig:clusters}
\end{figure*}

Table VII presents cluster characteristics.

\begin{table}[!t]
\caption{Cluster Characteristics}
\label{tab:clusters}
\centering
\resizebox{\columnwidth}{!}{%

\begin{tabular}{lccc}
\toprule
Cluster & N & Mean Overall Rating & Std Dev \\
\midrule
Cluster 0 (Defensive) & 4,335 & 64.3 & 6.2 \\
Cluster 1 (Offensive) & 5,334 & 69.0 & 5.3 \\
\bottomrule
\end{tabular}}
\end{table}

Clusters show different mean Overall Ratings (64.3 vs 69.0). This likely reflects that defensive specialists receive lower overall ratings in this evaluation system, which emphasizes offensive contributions. The key finding is that residual-space clustering (PC1 removed) identifies defensive-oriented versus offensive-oriented profiles with high stability (ARI = 0.97), though separation is modest (silhouette = 0.25), consistent with continuous variation.

Interpretation requires caveats. First, modest silhouette indicates continuous multidimensional space rather than discrete types. Second, high ARI indicates stable partitioning despite continuous structure. Third, cluster mean ratings differ, so clusters do not represent "similar overall quality with different profiles"; rather, they represent tendencies along a defensive-offensive continuum that correlate with overall ratings. This exploratory analysis complements main findings of four meaningful dimensions beyond PC1.

\section{Flexible Prediction Benchmark}

In this appendix, we report a Random Forest benchmark (R squared = 0.947, 5-fold CV). This serves only as an upper bound on recoverable information given the mechanical construction of Overall Rating, not as a substantive model. It confirms that the multidimensional attribute set contains substantial structure beyond what linear models capture.

\end{document}